R package for producing beamplots as a preferred alternative to the h index when assessing single researchers (based on downloads from Web of Science)

Robin Haunschild*, Lutz Bornmann** & Jonathan Adams***

* Max Planck Institute for Solid State Research

Heisenbergstraße 1,

70569 Stuttgart, Germany

E-mail: r.haunschild@fkf.mpg.de

** Administrative Headquarters of the Max Planck Society

Division for Science and Innovation Studies

Hofgartenstr. 8,

80539 Munich, Germany.

E-mail: bornmann@gv.mpg.de

*** ISI Clarivate Analytics, and,

The Policy Institute at King's,

King's College London,

22 Kingsway,

London WC2B 6LE, UK.

E-mail: jonathan.adams@kcl.ac.uk

**Abstract**

We propose the use of beamplots – which can be produced by using the R package BibPlots and WoS downloads – as a preferred alternative to h index values for assessing single researchers.

Since its introduction (Hirsch, 2005) the h index has received considerable attention and use in research evaluation. An h index can be calculated for any set of papers in Web of Science (Clarivate Analytics) and Scopus (Elsevier) for which citation data are also available. In a recent report from Clarivate Analytics (Adams, McVeigh, Pendlebury et al., 2019), attention was drawn to the use of beamplots, based on methodology described by Bornmann and Haunschild (2018). The authors recommended beamplots as an assessment tool. They argued – based on the work of Bornmann and Marx (2014a, 2014b) – that beamplots are advantageous because they combine the number of papers and their citation impact, as does the h index, but they do not reduce performance information to a single and somewhat arbitrary number. Adams et al. (2019) note that it is questionable to report multifaceted bibliometric performance in only one number. We would also note that no convincing reason exists why papers with h citations and not $h^2$ or $h*2$ citations should be counted in an index.

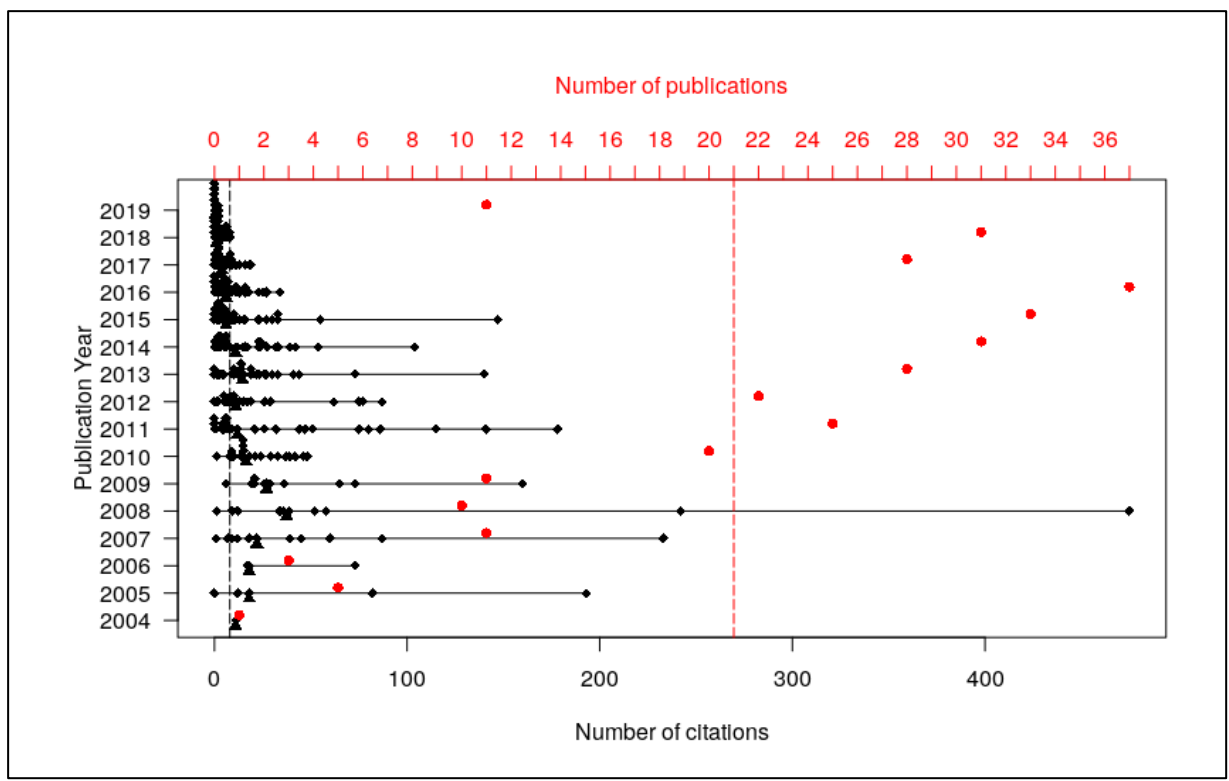

Figure 1. Beamplot based on publication and citation data from the Web of Science

Figure 1 is a beamplot for the publication set of a single researcher. The lower x-axis shows the number of citations that the papers received and the y-axis distributes these papers across the publication years. Each of the researcher's papers is shown, visualized with a black diamond. The lines (the 'beams') for a publication year visualize the range of citation counts for the papers published in that year. The black triangles below the lines indicate the median citation count of the papers published in that year. The dotted black line is the median number of citations across all years. The upper x-axis shows the count of papers in each year. The red circles show the number of publications published in a year, and the dotted red line is the median number of publications per year.

A beamplot is not only a simple yet comprehensive and informative presentation of a researcher's record. It also contains further information that would be valuable for the responsible assessment of the performance of individual researchers: year of first publication; interruptions in scientific output; and career progression. Productivity is revealed both for each single year and across all years. Citation impact is displayed for each individual paper, for the papers published in every year, and for all papers published by the researcher. It has to be kept in mind that citations need time to accumulate. It is to be expected that younger publications have lower citation counts than older papers. Such a pattern does not imply a reduced citation impact of the researcher over time. A rather simple weighting of the citation counts with the publication age can help to mitigate this disadvantage and any confusion that might arise. Most publications gather most of their citations within the first ten years after publication. Therefore, we propose a simplistic but wholly transparent linear weighting factor depending on the difference between the year until that citations are counted (i.e., the current calendar year in the case of WoS downloads) and the publication year. A weighting factor of 1 is used for a difference of 0, 1/2 for a difference of 1, …, and 1/11 for differences of ten or more. Such an option is also included in the R package BibPlots.

To make this analytical methodology more readily available to other researchers, we have expanded the functionality of the R package BibPlots, which can be used to generate the beamplot for a publication set downloaded from the Web of Science (in the "Other File Format → Tab-delimited (Win, UTF-8)" format). More detailed information about how to use the R package can be found at https://cran.r-project.org/web/packages/BibPlots/index.html.

We recommend the use of beamplots as a significant and preferred alternative to the h index whenever the assessment of single researchers is required. The beamplot in Figure 1 uses absolute citation counts. This can be justified when researchers are to be compared which are working in the same (or at least similar) fields and are of the same (or at least similar) scientific age. For the comparison of researchers from different fields, or where an individual has a publication record that spreads across multiple and diverse fields, field-normalized scores (e.g., citation percentiles, applied within a single year/subject category dataset) should be visualized instead of citation counts, as proposed by Bornmann and Marx (2014a, 2014b), Bornmann and Haunschild (2018) and Adams et al. (2019). When researchers in similar fields but of different age are to be compared, an alternative to the beamplot in Figure 1 can be used. Each citation count is divided by the age of the respective publications (see above).